\documentstyle[lanl,pslatex,rawfonts]{article}  
\begin{document}
\frompage{000} \topage{000}                                              
\def\rightmark{Hydrodynamic flow at RHIC} 
\def\leftmark{Peter F. Kolb}
\newcommand{\nc}{\newcommand}
\nc{\bb}{{\rm {\bf b}}}
\nc{\bs}{{\rm {\bf s}}}
\nc{\pt}{p_{\rm T}}
\nc{\mt}{m_{\rm T}}
\nc{\pL}{p_{\rm L}}
\nc{\ET}{E_{\rm T}}
\nc{\Nch}{N_{\rm ch}}
\nc{\Nc}{N_{\rm coll}}
\nc{\Np}{N_{\rm part}}
\nc{\Atanh}{{\rm Atanh}}
\nc{\Asinh}{{\rm Asinh}}
\nc{\Acosh}{{\rm Acosh}}
\nc{\se}{\section}
\nc{\suse}{\subsection}
\nc{\beq}[1]{\begin{equation}\label{#1}}
\nc{\eeq}{\end{equation}}
\nc{\bea}[1]{\begin{eqnarray}\label{#1}}
\nc{\eea}{\end{eqnarray}}
\nc{\bce}{\begin{center}}
\nc{\ece}{\end{center}}
\nc{\bit}{\begin{itemize}}
\nc{\eit}{\end{itemize}}
\nc{\bmp}{\begin{minipage}}
\nc{\emp}{\end{minipage}}
\newcommand{\gapp}{\raisebox{-.5ex}{$\stackrel{>}{\sim}$}}
\newcommand{\lapp}{\raisebox{-.5ex}{$\stackrel{<}{\sim}$}}
\newcommand{\av}[1]{\langle{#1}\rangle}
\nc{\la}{\langle}       
\nc{\lla}{\left \langle}
\nc{\ra}{\rangle}       
\nc{\rra}{\right \rangle}
\newcommand{\lda}{\langle\!\langle}       
\newcommand{\llda}{\left\langle\!\left\langle}
\newcommand{\rda}{\rangle\!\rangle}       
\newcommand{\rrda}{\right\rangle\!\right\rangle}
\newcommand{\eps}{{\varepsilon}}
\newcommand{\ds}{\displaystyle}
\newcommand{\half}{{\textstyle{1\over 2}}}
%
%
%
%
%
%
\title{Hydrodynamic flow at RHIC}
\authors{
{Peter F. Kolb$^{1,2,a}$}\\[2.812mm]
{\normalsize
\hspace*{-8pt}$^1$ 
Institut f\"ur Theoretische Physik \\ 
Universit\"at Regensburg\\
D-93040 Regensburg, Germany\\[0.2ex] 
\hspace*{-8pt}$^2$ 
Department of Physics\\
The Ohio State University\\ 
174 West 18th Avenue\\
Columbus, OH 43210, USA\\[0.2ex]
\hspace*{-8pt}$^a$ 
E-mail: pkolb@mps.ohio-state.edu\\
}}
%
%
%
%
%
%
\abstract{
We review the apparently hydrodynamic behaviour of 
low transverse momentum particles ($p_T \leq 1.5$ GeV/$c$) 
produced in central and semicentral ($b \leq 7$ fm)
heavy ion collisions at RHIC. 
We investigate the impact parameter dependence of
various observables, elaborating on radial and elliptic flow and 
particle multiplicities. 
We also discuss possible ambiguities in
the initialization of the hydrodynamic system
and present observables that should allow for their resolution.}

\maketitle
%
%
%
%
%
%
\section{Introduction and Motivation}\label{intro}

The first measurements at RHIC that systematically investigated 
the centrality dependence of an observable focused on elliptic flow 
(the anisotropic particle emission from the collision) 
\cite{STAR00}, followed by the centrality dependence 
of the absolute (charged) particle yield per unit of pseudorapidity \cite{PHENIX} 
and the produced transverse energy per unit of pseudorapidity \cite{PHENIX01}. 
Such systematic studies of the  influence of the collision  
centrality are of fundamental interest, as they represent a powerful tool to 
gain a detailed understanding of the collision dynamics:

Firstly, non-central collisions offer additional observables
due to their deformed, almond shaped overlap region, which 
can lead to angular dependencies (relative to the reaction plane)
of final state observables which do not appear in central collisions with azimuthal
symmetry \cite{Oll92}. 
Large anisotropies arise only if there is strong rescattering already 
in the first moments
( $\sim$ fm/$c$) of the collision, and (anisotropic) pressure gradients are 
building up, determining the subsequent evolution of the matter. Curiously 
the stronger forces in the direction of steepest pressure gradients lead to 
more transport of matter in those directions and thus eventually even out 
the differences between the radial gradients 
in the short and long direction of the initial
almond. Thus anisotropies that are observable in the
final state are built up early and in the hottest stages of 
the collision, as the cause of these
anisotropies disappears during the system's evolution (on a timescale of less 
than $\sim$ 4 fm/$c$ \cite{S97,ZGK99,KSH00} ).
In contrast to this self-quenching effect e.g. for elliptic flow, 
other dynamical quantities such as radial flow continue to grow until freeze-out
and carry information about the full expansion stage.
We explore the influence of the initial spatial anisotropies in terms of 
a hydrodynamic picture, which represents the limiting case of maximum 
response to the initially produced pressure gradients due to strong 
(infinite) rescattering already in the early stages of the expansion. Such an 
approach  was shown to be appropriate at RHIC energies \cite{KH3} and 
is  valuable to understand the global (macroscopic) characteristics of 
the expansion stage of an ultrarelativistic heavy ion collision.

Secondly, changing the centrality leads to a varying number of 
participating nucleons and a changing  size of the interaction region.
The amount of  energy deposited in the collision region as well 
as the energy density in the system will be largest in central 
collisions and decrease with increasing impact parameter. 
Thus by varying the centrality, one is able to scan the initial energy density
and in this fashion can measure excititation functions even without varying the beam
energy. It is crucial however to disentangle such 'centrality excitation functions' 
from the geometric effects introduced by the varying excentricity of the system.
In this spirit we investigate the centrality dependence of
particle production per participating nucleon pair and transverse energy 
carried by the emitted hadrons to learn about soft and hard scattering 
contributions in the initial processes.
 
In section \ref{hydro}, we introduce the underlying assumptions of hydrodynamic
models and focus especially on different initialization scenarios.
We present our results  and comparisons to experimental data
in Section 3, which covers a discussion of particle spectra and radial flow,
elliptic flow, multiplicities and transverse energy. 
Section 4 contains a brief summary. 
In the Appendix we study the effect of boost invariance on the 
pseudorapidity dependence of multiplicities and elliptic flow.
This helps to understand the corresponding shapes of the recently 
presented experimental data \cite{PHOBOStalks} around midrapidity.


\section{Hydrodynamics and Initialization}\label{hydro}  

Hydrodynamics is a macroscopic approach to describe the dynamical evolution
of the expansion stage of a heavy ion collision. 
It is a phenomenological model that,
by describing the evolution of thermodynamic fields like 
energy density, pressure, temperature and flow fields, 
circumvents the necessity of introducing unknown microscopic 
parameters (e.g. in-medium cross sections or string tensions) as required 
for microscopic descriptions of such systems.

In the hydrodynamic description the nuclear equation of state enters 
the model quite naturally 
as the connection of pressure or temperature to energy and particle density.
However it is not at all obvious that such an approach is feasible at all, as a 
thermodynamic treatment requires a 'large', 'macroscopic'
system in local thermal equilibrium and an adiabatic expansion stage.
But the good agreement of hydrodynamic simulations 
and experimental data from RHIC in fact point towards such  rapid thermalization
followed by a  hydrodynamic expansion.

The effective treatment however lacks a physical understanding of the 
underlying microscopic processes and the early equilibration time
as well as an explanation of the
obviously very large rescattering rates. 
It must then eventually be supplemented by a microscopic kinetic treatment 
to check its validity.
Finally a
hydrodynamic approach is only valid for a certain timespan of the expansion. 
We therefore have to introduce assumptions about initialization and 
freeze-out conditions \cite{KH3,KH2ET01}.

Existing microscopic models on the other hand so far lack 
rescattering mechanisms which are strong
enough to explain the large observed anisotropies (see e.g. \cite{BS00}), or 
the need to assume unrealistically large cross-sections \cite{MG01}. Only when
coupled to a hydrodynamic initial stage that is able to generate sufficient flow
anisotropies before the system enters the hadronic rescattering phase, the
large observed anisotropies can be recovered \cite{TLS01}. 
Recently however
progress in purely microscopic models was reported by introducing 
multi-Pomeron exchanges in quark gluon string models \cite{ZFBF01}.

\subsection{Relativistic hydrodynamics and equation of state}\label{hydro2}

Adiabatic expansion of matter is described by the hydrodynamic equations for the
conservation of energy, momentum and baryon number.
In relativistic form they read 
\beq{econs}
	\partial_\mu T^{\mu \nu}=0,\quad \partial_\mu j^\mu =0\,,
\eeq
with the energy-momentum tensor and the baryon current 
\beq{etensor}
	T^{\mu \nu}(x)=(e(x)+p(x))\,u^\mu(x)u^\nu(x)-g^{\mu\nu}p(x)\,,\quad
        j^\mu(x)=n(x)u^\mu(x) \;.
\eeq
These equations for the space-time evolution of the physical fields
are closed by an equation of state (EoS) 
relating energy and baryon density to the pressure and temperature. The EoS 
for this study contains a sharp first order phase transition, which connects a 
hadronic resonance gas at low energy densities ($e < 0.45$ GeV/fm$^3$) 
to the hard equation of state of an ideal ultrarelativistic gas 
(modeling a quark gluon plasma phase) in the high energy 
density region of the phase space diagram ($e>1.6$ GeV/fm$^3$). 
Further details on the equation of 
state and its construction can be found in \cite{KSH00}, which also includes 
a discussion of the influence of the phase transition and details of the 
equation of state on final state observables.

To reduce computational costs we analytically implement boost-invariance 
in longitudinal direction. The shape of the measured $dN/d\eta$ distribution
\cite{QM01} indicates that this is well satisfied around midrapidity at RHIC 
energies (see Appendix). Fully three-dimensional calculations exist for SPS
energies \cite{HN} and are under development for RHIC energies.
\subsection{Initialization}\label{initial}
 
The linear scaling of particle production with the number of wounded nucleons 
as observed at SPS energies \cite{WA98} indicates that the first 
scattering processes underlying such collisions 
are soft or non-perturbative, and that it is the number  
of wounded nucleons that governs particle production.
We assume this to hold locally in the transverse plane and the number of 
produced particles to be proportional to the number of participants 
per unit area. 
For a collision with impact parameter $\bb$ this density 
at a point $\bs$ in the transverse plane 
is given by
 \bea{init}
 n_{\rm WN}(\bs;\bb) =
 T_A\bigl(\bs{+}\half\bb\bigr)
      \Bigl[1-\Bigl(1-{\sigma T_B\bigl(\bs{-}\half\bb\bigr)
                       \over B}\Bigr)^B \Bigr] 
   + T_B\bigl(\bs{-}\half\bb\bigr)
     \Bigl[1-\Bigl(1-{\sigma T_A\bigl(\bs{+}\half\bb)
                      \over A}\Bigr)^A \Bigr]\,,
\nonumber
 \eea
where we have introduced the nuclear thickness function
 \bea{TA}
   T_A(\bs)=\int_{-\infty}^{+\infty} dz\, \rho_A(\bs,z) \,,
\nonumber
\eea
with $\rho_A(\bs,z)$ parametrizing the nuclear density profile (i.e. a
Woods Saxon profile with appropriate parameters for a nucleus with mass number A).

On the other hand one expects the particle production from hard
collisions to dominate at high energies. This perturbative
particle production scales with the number of collisions, which is
given per unit area in the transverse plane by
\bea{TAB}
   n_{\rm BC}(\bs;\bb) = \sigma\,
   T_A\left(\bs{+}\half\bb\right) T_B\left(\bs{-}\half\bb\right)\,.
\nonumber
 \eea

In the following we study the results obtained using 
five different parametrizations of the initial state. 
We assume either energy or entropy density to be proportional 
to $n_{\rm WN}$  (parametrizations labeled eWN and sWN, respectively) 
or to  $n_{\rm BC}$ (labeled eBC and sBC, respectively).
In addition we use an initialization resulting from a saturation
model calculation which limits the growth of the gluonic cross-section 
in the transverse plane on the basis of geometrical arguments \cite{EKRT00},
and label these results by 'sat'. More details on the initialization 
models can be found in \cite{KH2ET01,EKRT00}. 


\section{Experimental Observables}\label{observe}
 
The initialization fields for energy and baryon number density
are obtained by a straightforward extrapolation from earlier simulations 
where we tuned  them to fit particle spectra 
resulting from the most central Pb+Pb collisions at the highest available SPS
beam energies \cite{KSH99}. 
For RHIC energies, we readjust only one parameter (the initial energy density
at the origin in $b=0$ collisions) to reproduce the 
final particle  multiplicity observed in central collisions at RHIC \cite{PHOBOS}.
(The equilibration time is scaled down so that its product with the 
maximum temperature is the same for RHIC and SPS systems).
In our hydrodynamic simulations, this results in a maximum energy density  
$e_0=$23 GeV/fm$^3$ at an equilibration time $\tau_0=0.6$ fm/$c$
(compared to 9 GeV/fm$^3$ at 0.8 fm/$c$ for the earlier SPS simulations) 
The new initialization leads to a mean energy density of 
$e_{\rm mean}= 4.8$ GeV/fm$^3$ at $\tau=1.0$ fm/$c$, for which   
the PHENIX collaboration estimates approximately 5.0 GeV/fm$^3$ \cite{PHENIX01}.

\subsection{Spectra and radial flow}
\begin{figure}[htb]
\bce
\begin{minipage}[b]{6.1cm}
           \epsfxsize 6.cm      \epsfbox{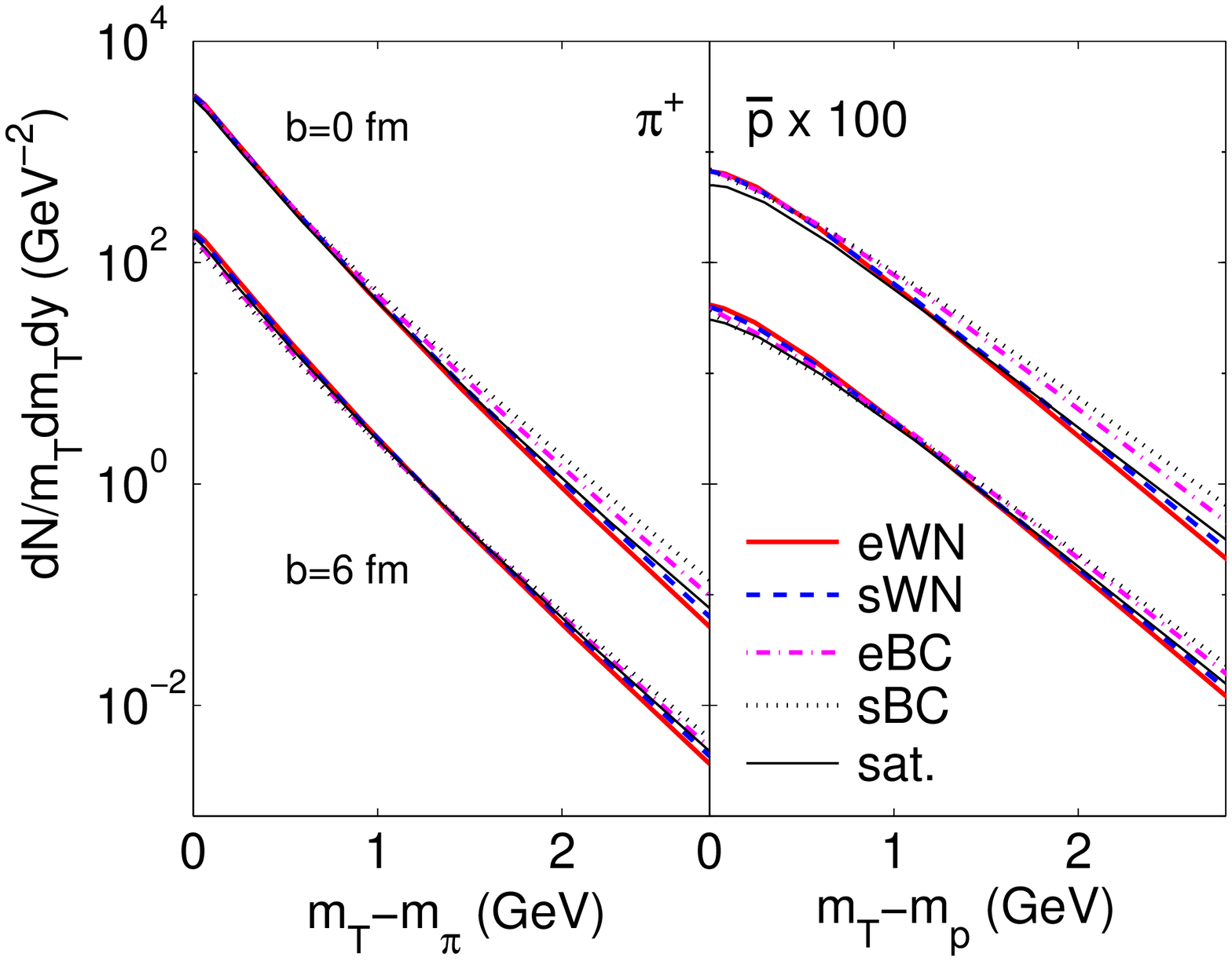}
\caption[]{Transverse mass spectra of pions and antiprotons for central
           and semicentral collisions (the latter scaled by a factor 0.1) 
           as resulting from the different initializations.}
\label{spectra}
\end{minipage}
	   \hspace{.2cm} 
\begin{minipage}[b]{6.1cm}
           \epsfxsize 5.6cm      \epsfbox{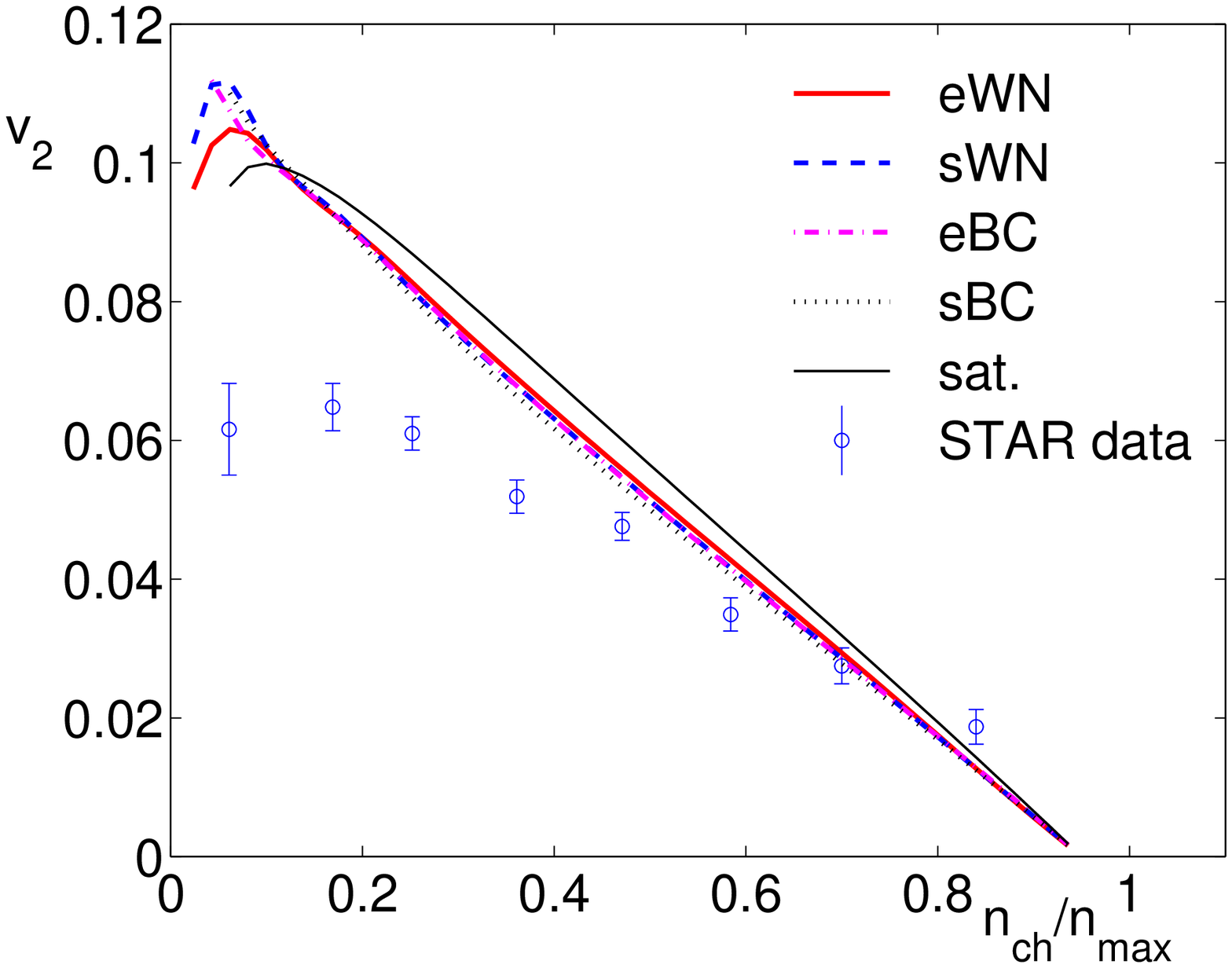}
\caption[]{Elliptic flow of charged hadrons as function of centrality, 
           given by the number of produced particles
           for the different initializations together with
           experimental data \cite{STAR00}.}
\label{v2nch}
\end{minipage}
\ece
\end{figure}

The larger energy densities and pressure gradients in calculations for RHIC 
lead to stronger transverse expansion, which is reflected in flatter 
transverse mass spectra compared to SPS results
(Fig. \ref{spectra} in \cite{HKHRV}). 
This is in 
good quantitative agreement with preliminary spectra from the RHIC 
experiments \cite{QM01}. In the hydrodynamic simulations, the average radial 
flow velocity at freeze-out increases 
from 0.45 $c$ at maximum SPS-energies to 0.55 $c$ at
$\sqrt{s_{NN}}=130$ GeV. The influence of the initialization on the slopes
of the particle spectra is weak, but one observes that the spectra are getting
flatter in the order eWN, sWN, eBC, sBC with the results for the saturation model 
somewhere in between sWN and eBC. This reflects the harder 
initializations of the binary collision models, which results in steeper 
initial pressure gradients and
larger driving forces.

\subsection{Elliptic flow}

\begin{figure}[htb]
\bce
\begin{minipage}[t]{6.1cm}
           \epsfxsize 6.cm      \epsfbox{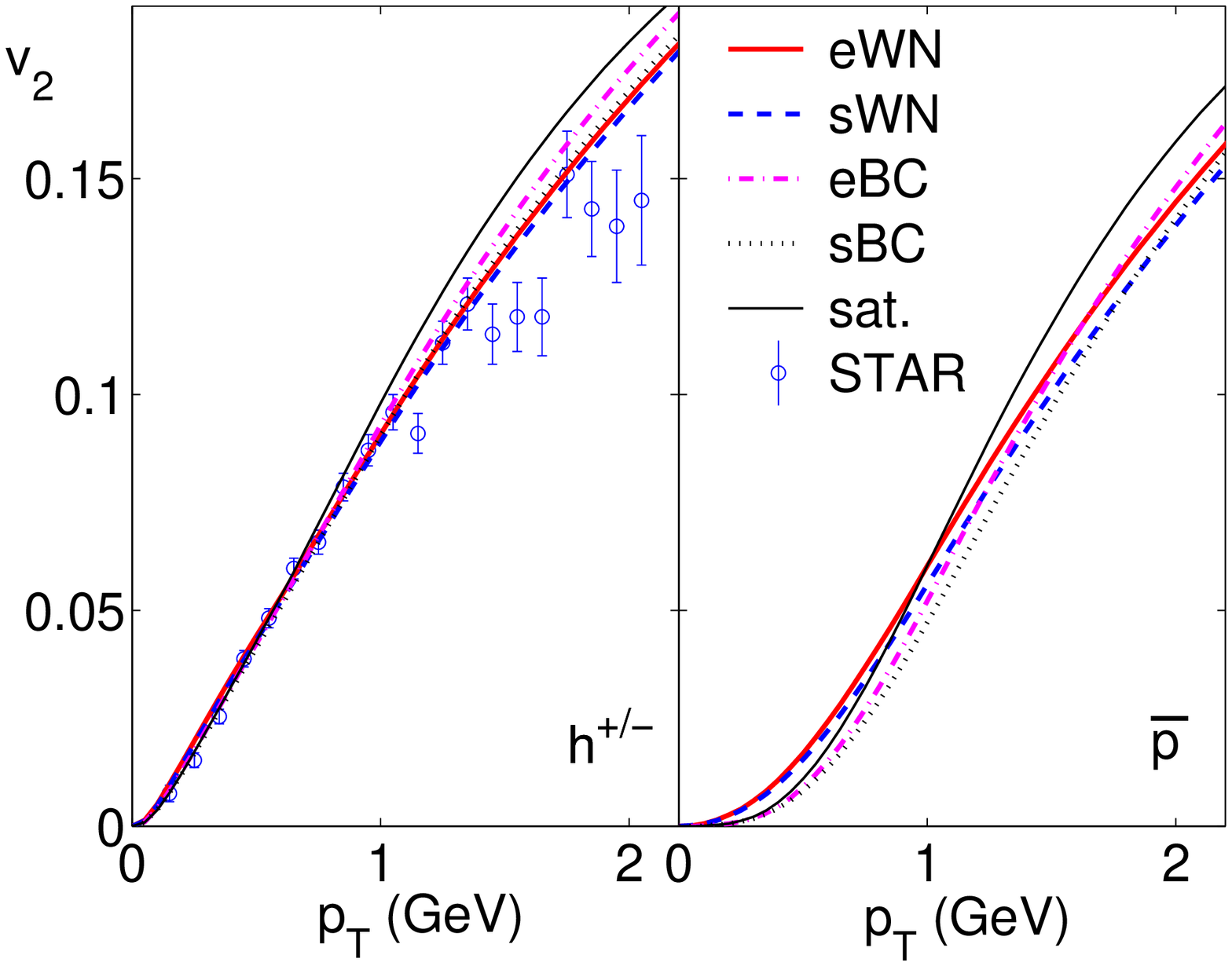}
\caption[]{Elliptic flow in minimum bias configuration for charged hadrons together
with the experimental data (left) and antiprotons (right).}
\label{v2pt}
\end{minipage}
	   \hspace{.2cm} 
\begin{minipage}[t]{6.1cm}
           \epsfxsize 6cm      \epsfbox{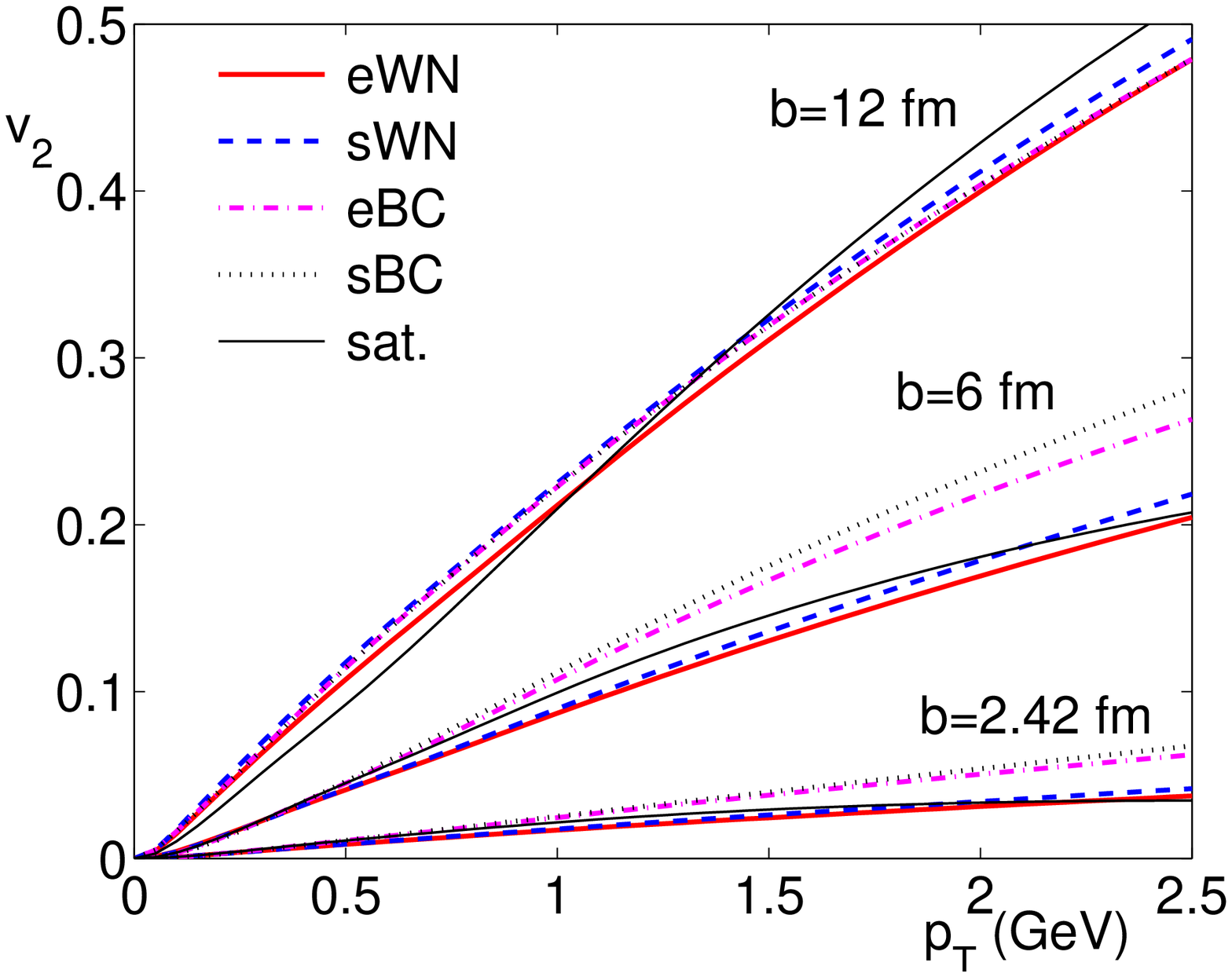}
\caption[]{Elliptic flow of charged hadrons as function of transverse momentum 
for fixed impact parameters.}
\label{v2ptbins}
\end{minipage}
\ece
\end{figure}

The quantitative agreement of hydrodynamic simulations with the measured 
data for elliptic flow, both for the momentum integrated and the minimum
bias averaged differential elliptic flow 
(for not too large impact parameters and transverse momenta), 
was pointed out and discussed in an earlier work 
\cite{KH3} and is reproduced in Figs. \ref{v2nch} and \ref{v2pt}. 
Here we investigate the influence of the different initialization 
scenarios. 

Fig. \ref{v2nch} shows the momentum integrated elliptic 
flow as a function of centrality, characterized by the particle yield 
at midrapidity. There is good agreement with the experimental data for central 
to semicentral collisions (high to intermediate values of $n_{\rm ch}/n_{\rm max}$)
independently of the underlying initialization (except of the saturated
initialization, giving larger anisotropies than the others). Also 
 Fig. \ref{v2pt}, which shows minimum bias averaged differential elliptic flow 
$v_2(p_T)$, reveals that the results for charged hadrons are rather independent of the 
underlying initialization. Deviations can only be seen at intermediate to high $p_T$, 
and then again especially for the saturation model. The analysis of charged hadrons is
dominated by pions due to their large abundancies. Analyzing heavier particles 
independently, as e.g. done for antiprotons in the right panel of 
Fig. \ref{v2pt}, shows that elliptic flow
of heavier particles is sensitive on the details of the initialization. 
Fig. \ref{v2ptbins} shows
differential elliptic flow for specific impact parameters, 
without avaraging over them to yield minimum bias results.
Also here a sensitivity on the initialization is seen, especially for
semicentral collisions. This sensitivity is lost when averaging over impact parameters
which goes ahead with weighting over the resulting particles, 
as seen in Fig. \ref{v2pt}. The centrality dependence of particle production in the
different models is thus crucial and therefore studied in the next subsection.

It is truly astonishing that the experimental data and the 
hydrodynamic results coincide up to impact parameters of about 7 fm and $p_T$
of about 2 GeV.
Out of all models applied to relativistic heavy ion collisions, 
hydrodynamics exhibits the
strongest (namely infinite) rescattering, and thus leads to the strongest mapping of 
initial coordinate anisotropies to final momentum space anisotropies. 
Hydrodynamics thus gives the upper limit of possible $v_2$. 
That the data reach up to this limit is remarkable!

\subsection{Multiplicities and transverse energy} 

\begin{figure}[htb]
\bce
\begin{minipage}[t]{12.9cm}
\begin{minipage}[t]{6.1cm}
\bce
     \epsfxsize 6.cm      \epsfbox{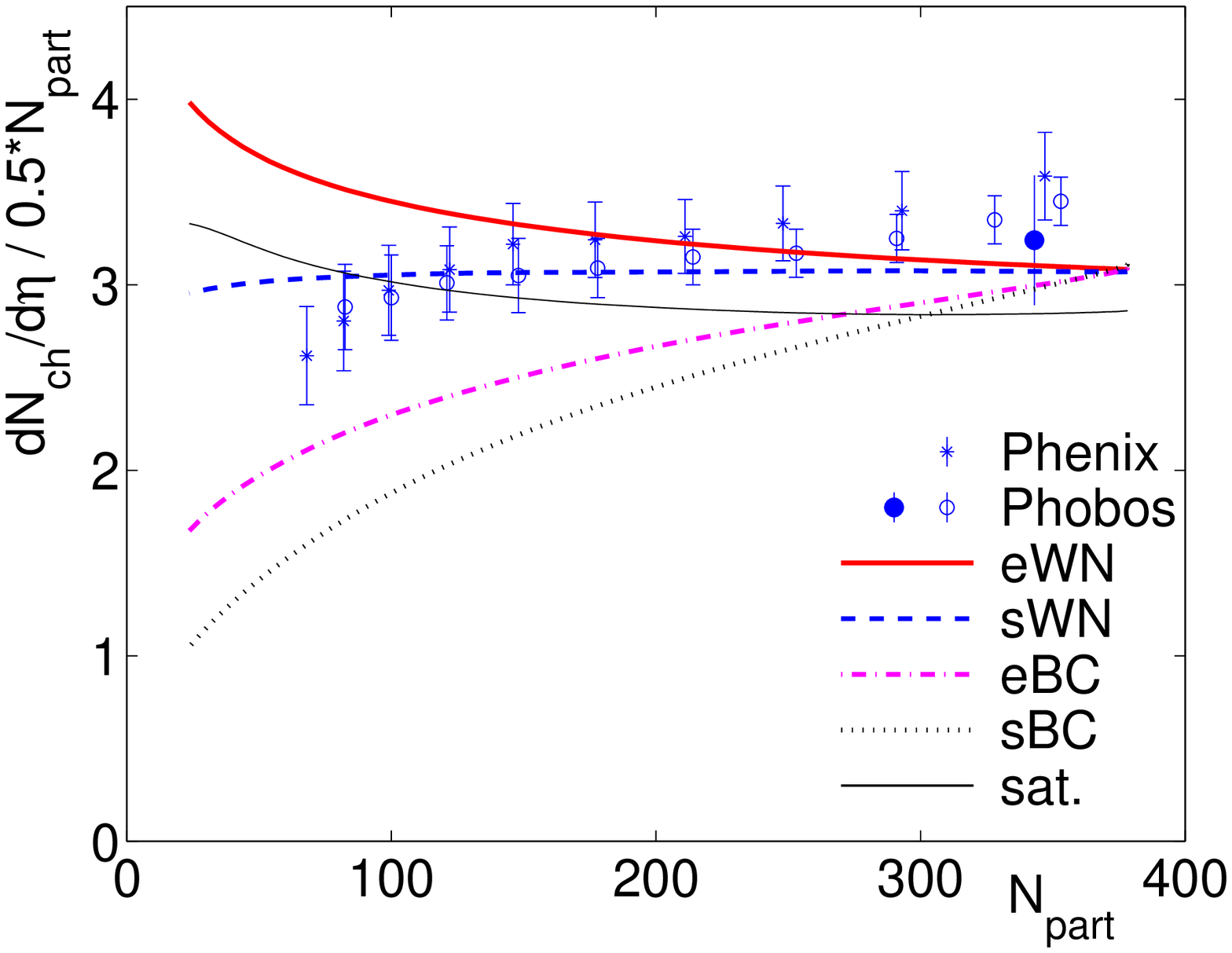}
\ece
\end{minipage}
\begin{minipage}[t]{6.1cm}
\bce
           \epsfxsize 6.cm      \epsfbox{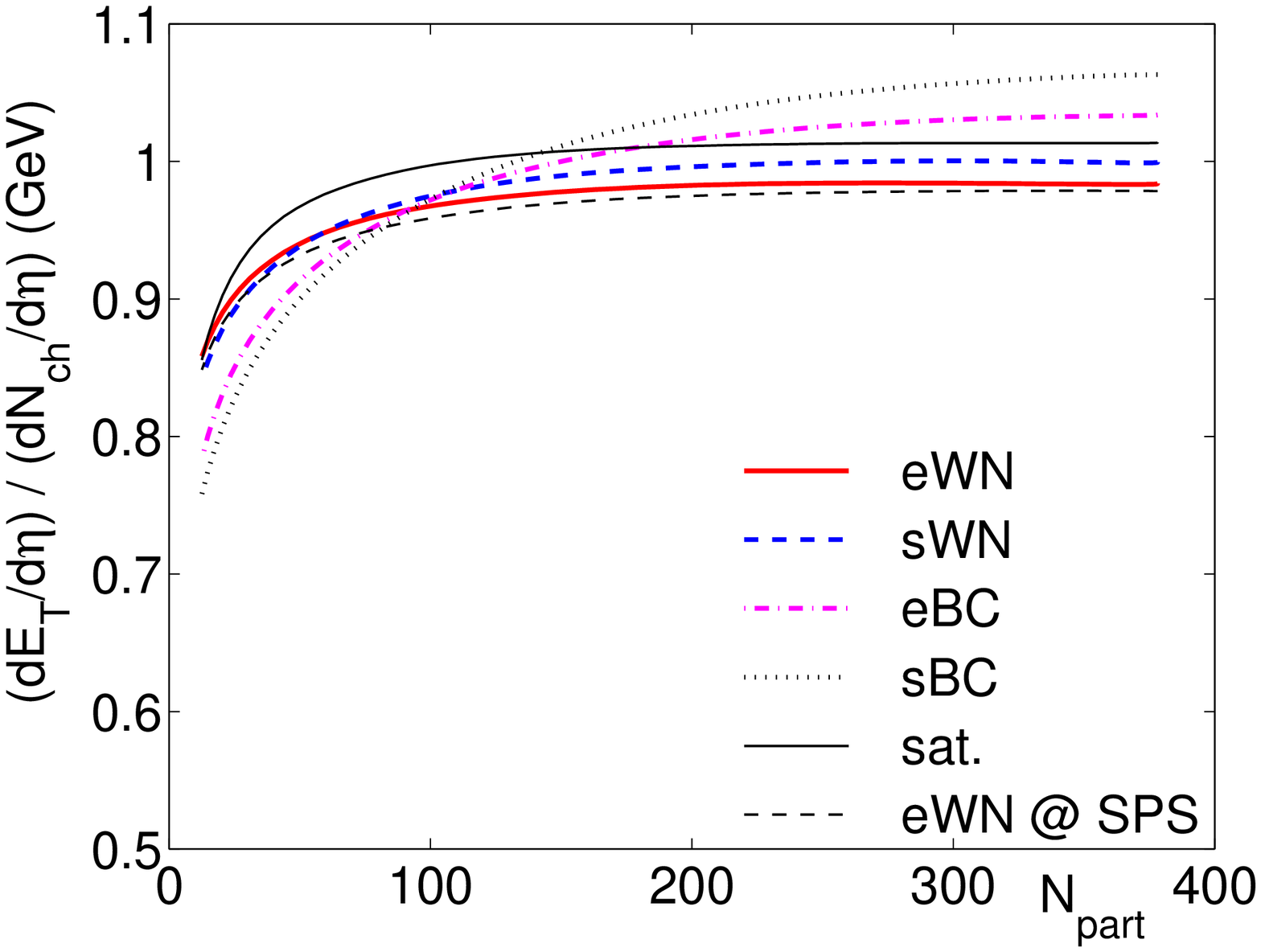}
\ece
\end{minipage}
\caption[]{Charged particle yield per participating nucleon pair 
           (left) and 
           transverse energy per produced particle (right) as functions 
           of participating nucleons.}
\end{minipage}
\label{multplots}
\ece
\end{figure}

Contrary to the minimum bias elliptic flow analysis, the centrality dependence of
particle production is quite sensitive on the details of initialization. 
The left plot in Fig. \ref{multplots} shows the number of 
produced charged hadrons per participating
nucleon pair as a function of participants from our simulations together with 
experimental results \cite{PHENIX,QM01,PHOBOS}.
The experimental data disfavor the soft initialization 'eWN' and the
saturation model. The other initializations show a tendency similar to the data,
i.e. rising particle production for more central collisions. The data would be 
best described by a combination \cite{KN00} of the two extremes for 
hard and soft scattering contributionsc to particle production 
that we have studied here.
This is under current study within our hydrodynamic approach.

The results for the transverse energy per produced particle are shown in the righthand
side plot of Fig. \ref{multplots}. Data from the SPS \cite{WA98} shows qualitative 
agreement with the shape of the curve from hydrodynamics initialized by a
wounded nucleon Ansatz -- a  saturating transverse energy per 
particle with a wide plateau from semicentral to central collisions.
However very recent experimental data at RHIC \cite{PHENIX01} show the same shape,
disfavoring the hard initializations of our model, which do not follow this trend. 
A definite statement on that 
seemingly contradictory behaviour of the centrality dependence of particle 
production and transverese energy carried per particle 
has to await a more careful
theoretical analysis which is under way.

\section{Conclusions}\label{concl}
We have reviewed the experimental and theoretical evidence for early equilibration
and subsequent hydrodynamic evolution of matter created in heavy ion collisions at 
RHIC. The large radial and anisotropic flow is most easily explained under the 
assumption of strong pressure gradients driving the system's expansion. The 
energy densities at the equilibration timescale reach far beyond the critical 
energy density.
Microscopic simulations with standard scattering cross sections fail to describe
flow observables due to a lack of sufficiently large rescattering.

We further investigated a possible influence of the initialization on the observables.
We found that no ambiguity arises for the analysis of the published results on 
centrality dependence of momentum integrated elliptic flow and transverse momentum 
dependence of elliptic flow in minimum bias configurations. On the other hand, we 
discussed observables which would allow for a distinction of the initialization,
that is minimum bias elliptic flow for heavier particles (e.g. antiprotons), 
momentum dependence of elliptic flow for  fixed intermediate impact parameters
and the transverse energy per produced particle.

\section*{Appendix}\label{app}
We briefly discuss the influence of the rapidity ($y$) $\leftrightarrow$ pseudorapidity
($\eta$) transformation on the pseudorapidity dependence of particle yield 
and elliptic flow
under assumption of boost-invariance along the beam axis. This will lead to a 
quantitative understanding of the experimental data recently presented by the
PHOBOS collaboration \cite{PHOBOStalks}.

We start out with the definitions for the rapidity 
$y:=\Atanh\left({p_z}/{E}\right)$ and the pseudorapidity
$\eta:=\Atanh \left( {p_z}/{p}\right)$. From these definitions follows the 
connection of the differentials $dy= \frac{p}{E} \cdot d\eta$ which is 
the focus of interest. 

For this instructive (quantitative) excursion we make very simple 
model assumptions. We stricly assume boost invariance not only around 
midrapitity, but assume that all quantities are independent of $y$, 
no matter how large $y$ is. For the spectra we use the simple exponentials
\[  
\frac{dN}{p_Tdp_Tdy}(p_T)= N e^{-\sqrt{m^2+p_T^2}/T}
\]
where we use for these case studies simply $T=190$ MeV and $m=m_{\pi}=140$ MeV.

Boost invariant, i.e.  $y$ independent spectra transform due to the Jacobian to 
$\eta$ dependent spectra according to 
\[
 \frac{dN}{p_Tdp_Td\eta}(p_T,\eta)= \frac{p_T \cosh \eta}{\sqrt{m^2+p_T^2 \cosh ^2 \eta}} 
                                    \frac{dN}{p_Tdp_Tdy}(p_T)\,.
\]
This distribution gives thus smaller values than the original one. 
The suppression is largest for low $p_T$ and vanishes for high $p_T$. 
Also the reduction is larger the larger the mass is, and the smaller $\eta$ 
(i.e. around midrapidity).
For large $\eta$ the distribution approach each other.

\begin{figure}
  \begin{center}
\begin{minipage}[t]{12.9cm}
  \begin{center}
\begin{minipage}[t]{6.1cm}
\bce
    \epsfxsize 6.cm \epsfbox{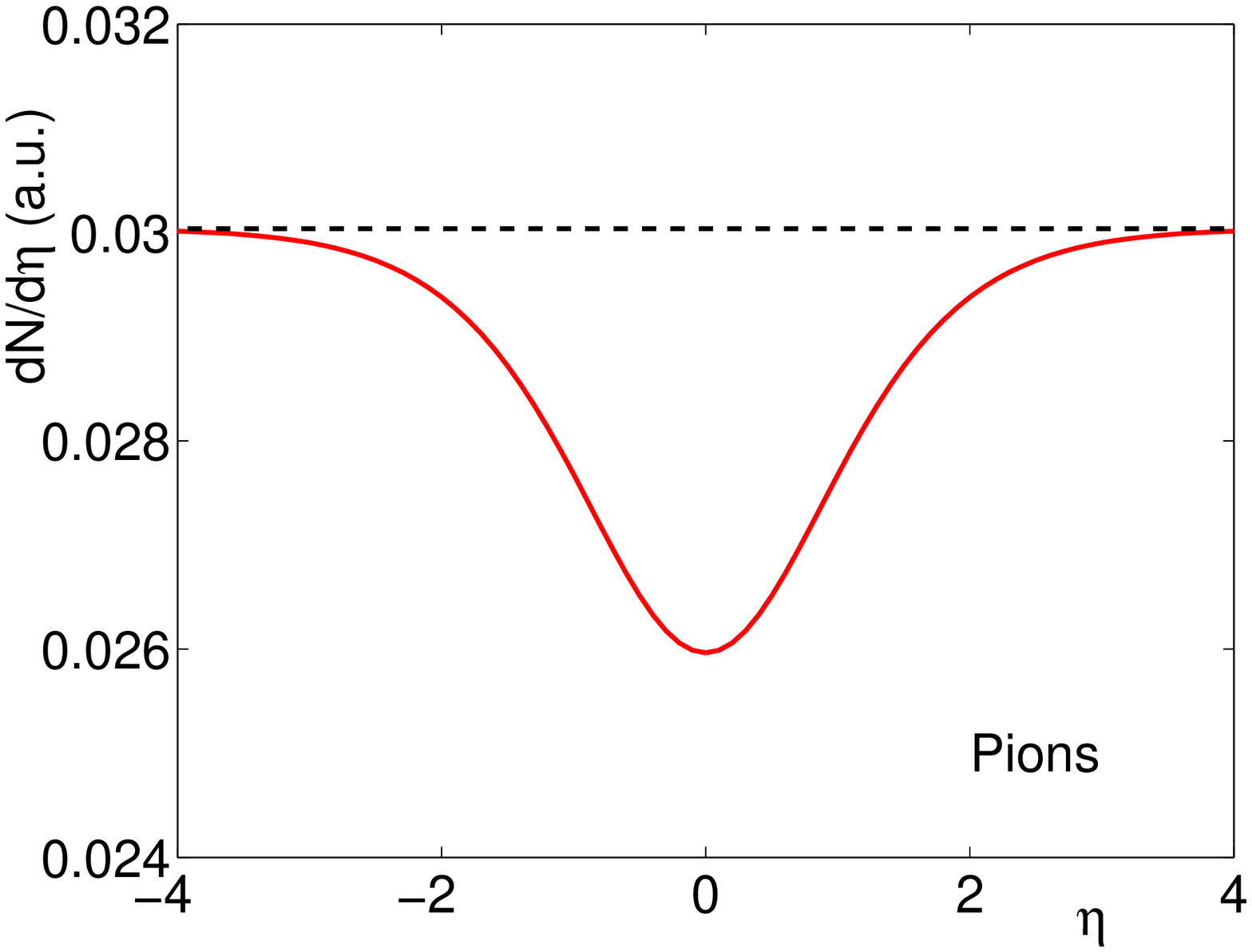}
\ece
\end{minipage}
\begin{minipage}[t]{6.1cm}
\bce
    \epsfxsize 6.cm \epsfbox{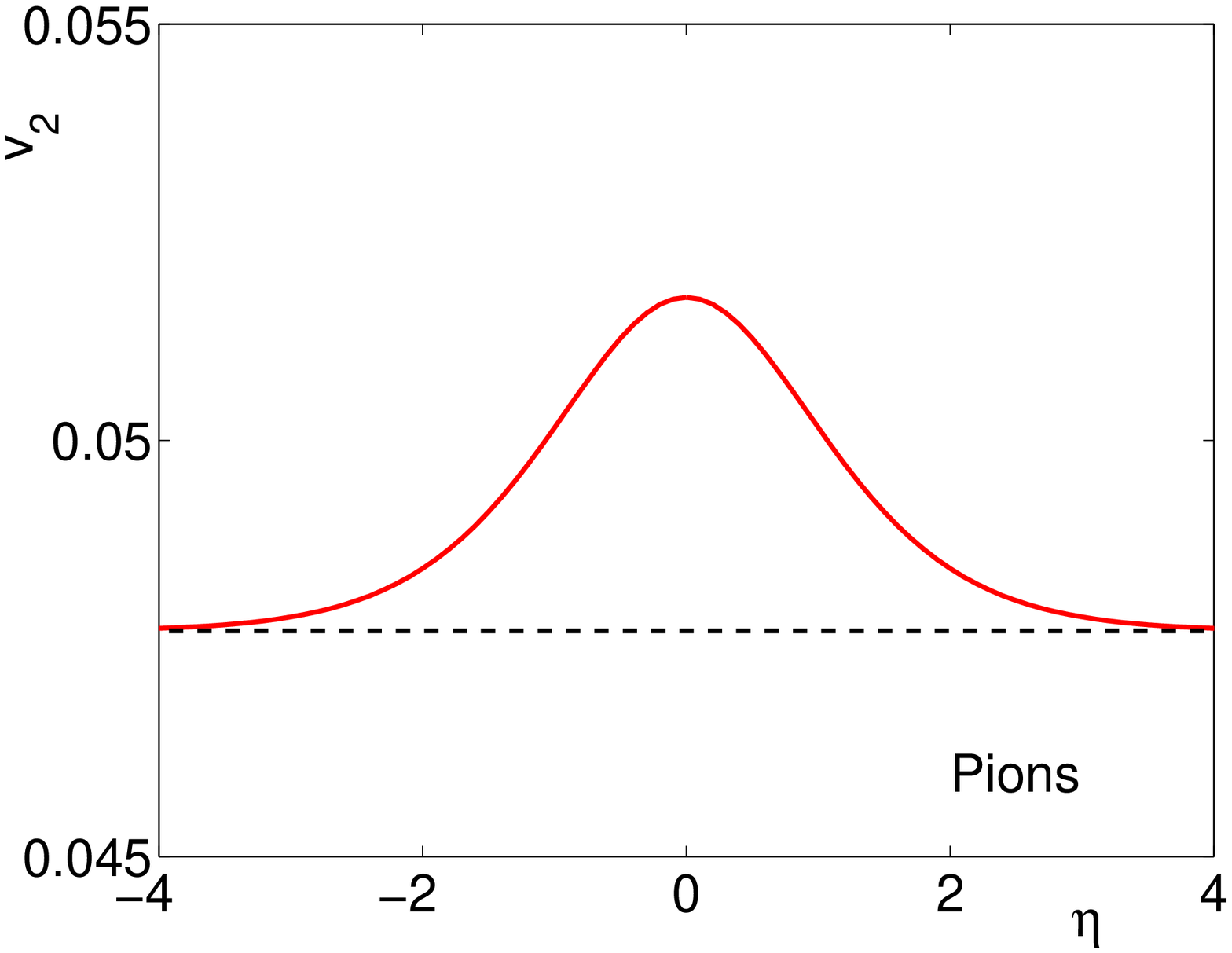}
\ece
\end{minipage}
  \end{center}
\end{minipage}
  \end{center}
\caption[]{Transformation of boost-invariant quantities 
  (independent of rapditity $y$) to pseudorapidity. The left plot shows the effect
  of the Jacobian of the transformation on the particle spectra, the right plot
  the influence on the elliptic flow coefficient $v_2$.}
\label{trafoplots}
\end{figure}

From this it is obvious that the $p_T$ integrated spectra lead
to smaller values of  $dN/d\eta$ around midrapidity than far away, where it 
approaches the constant value of $dN/dy$ (left panel of Fig. \ref{trafoplots}). 
In reality 
boost invariance must break down at some rapidity, and the particle yield
drops to zero. However the dip observed  in the $dN/d\eta$ distribution 
around midrapidity is thus well expected from a boost-invariant source, and 
we find in fact quantitative agreement with the preliminary STAR-data \cite{QM01}
from $\eta=-0.5$ to 0.5.

Rapidity dependent elliptic flow is defined 
as $v_2(p_T,b,y)=\langle \cos(2\phi) \rangle$ 
and the same for $v_2(p_T,b,\eta)$ where the averages are taken with respect
to the $dy$ respectively $d\eta$ distributions with a fixed impact 
parameter $b$. From the definitions one finds easily that 
$v_2(p_T,b,\eta)=v_2(p_T,b,y)=v_2(p_T,b,y=0)$ and therefore also 
$v_2(p_T,b,\eta)$ is independent of the pseudorapidity.
However for the total elliptic flow, when integrating over the 
transverse momentum, the pseudorapidity dependence will come into play, due to 
the different shapes of the spectra:
\[ v_2(\eta)=\frac{ \int  dp_T \,p_T v_2(p_T) \frac{dN}{p_T dp_T d\eta}(\eta)}
                  {\int dp_T \, p_T \frac{dN}{p_T dp_T d\eta}(\eta)}\,.
\]

Consider the effect of this: $v_2(p_T)$ is a monotonically increasing
function. 
For large $\eta$ the weighting particle distributions will become identical
to the spectra $dN/p_T dp_T dy$ and therefore we will recover $v_2$ as 
weighted with the boost-invariant rapidity-distributions.
Now going to smaller pseudorapidities we are weighting the differential
elliptic flow with the suppcression of the low $p_T$ part. Thus higher 
$p_T$'s with larger $v_2(p_T)$ will get stressed and we will have larger
$p_T$-integrated
elliptic flow around mid(pseudo)rapidity, than away from it! We illustrate
this in Fig. \ref{trafoplots}, where we used for simplicity the linear relation 
$v_2(p_T,\eta)= p_T(\mbox{GeV})/8.5$ which is a first approximation for the
experimental results on 'minimum bias' collisions.

Thus, a boost invariant source would show a bump in $v_2(\eta)$ 
and a dip in the rapidity distribution $dN/d\eta$. In reality we also expect
that elliptic flow drops to zero far away from mid-rapidity
as in these regions the reaction dynamics does not allow for sufficient 
equilibration, the primary cause of elliptic flow in non-central collisions.
 
\section*{Acknowledgements}
I would like to thank the organizers for the stimulating workshop and especially 
acknowledge interesting and motivating discussions with J.~Aichelin, B.~Back, 
C.~Bertulani, R. Lacey, P.~Seyboth, and G.~Verde.

I have reported on work done in collaboration with K. Eskola, U. Heinz, 
P. Huovinen and K.~Tuominen. P. Huovinen and U. Heinz are thanked for help with
the manuscript.
This work was supported in parts by the Deutsche Forschungsgemeinschaft.


\vfill\eject
\end{document}